\documentclass[conference]{IEEEtran}
\usepackage{tikz}
\usepackage{amsmath}
\usepackage{amssymb}
\usepackage{amsfonts}
\usepackage{paralist}
\usepackage{xspace}
\usepackage{multirow}
\usepackage{listings}
\definecolor{lightlightgray}{RGB}{238,238,238}
\usepackage{hyperref}
\usepackage{url}

\ifCLASSOPTIONcompsoc
  \usepackage[nocompress]{cite}
\else
  \usepackage{cite}
\fi

\ifCLASSINFOpdf
\else
\fi
\begin{document}
%
\newcommand{\acmatrix}{Exploit Classification\xspace}
\newcommand{\acmodel}{Modeling\xspace}
\newcommand{\acplan}{AI Planning\xspace}
\newcommand{\name}{ALFA-Chains\xspace}
\newcommand{\balfachains}{\name}
\newcommand{\RCE}{RCE\xspace}
\newcommand{\coreblah}{Core Certified Exploit Library\xspace}
\newcommand{\networkTwo}{DMZ+LAN\xspace}
\newcommand{\networkTwenty}{20+6subs\xspace}
\newcommand{\networkTwoHundred}{200+6subs\xspace}
\newcommand{\purdueOne}{$\textnormal{Purdue}_1$}
\newcommand{\purdueTwo}{$\textnormal{Purdue}_2$}
\newcommand{\purdueThree}{$\textnormal{Purdue}_3$}
\newcommand{\synthetic}{Synthetic\xspace}
\newcommand{\hostzero}{H0\xspace}
\newcommand{\hostone}{H1\xspace}
\newcommand{\hosttwo}{H2\xspace}

\title{\Large \bf Hybrid Privilege Escalation and Remote Code Execution Exploit Chains}


\author{\IEEEauthorblockN{Miguel Tulla}
\IEEEauthorblockA{MIT\\
Cambridge, MA\\
Email: mtulla@mit.edu}
\and
\IEEEauthorblockN{Andrea Vignali}
\IEEEauthorblockA{University of Naples Federico II\\
Naples, Italy\\
Email: andrea.vignali@unina.it}
\and
\IEEEauthorblockN{Cristian Colon}
\IEEEauthorblockA{MIT\\
Cambridge, MA\\
Email: ccolon@mit.edu}
\and
\IEEEauthorblockN{Anahita Srinivasan}
\IEEEauthorblockA{MIT\\
Cambridge, MA\\
Email: anahi183@mit.edu}
\and
\IEEEauthorblockN{Giancarlo Sperlì}
\IEEEauthorblockA{University of Naples Federico II\\
Naples, Italy\\
Email: giancarlo.sperli@unina.it}
\and
\IEEEauthorblockN{Simon Pietro Romano}
\IEEEauthorblockA{University of Naples Federico II\\
Naples, Italy\\
Email: spromano@unina.it}
\and
\IEEEauthorblockN{Masataro Asai}
\IEEEauthorblockA{MIT-IBM Watson AI Lab\\
Cambridge, MA\\
Email: masataro.asai@ibm.com}
\and
\IEEEauthorblockN{Erik Hemberg}
\IEEEauthorblockA{MIT\\
Cambridge, MA\\
Email: hembergerik@csail.mit.edu}
\and
\IEEEauthorblockN{Una-May O'Reilly}
\IEEEauthorblockA{MIT\\
Cambridge, MA\\
Email: unamay@mit.edu}
}

\maketitle

\begin{abstract}

Research on exploit chains has predominantly focused on sequences involving only one type of exploit, for example, either escalating privileges on a single machine or executing remote code on a network. 
Notwithstanding, in modern networks, hybrid exploit chains are critical because of their linkable vulnerabilities.  
At the same time, developing hybrid exploit chains is challenging because it requires understanding the diverse and independent dependencies and outcomes of each exploit, which can result in an overwhelming number of possibilities.
We present hybrid chains encompassing privilege escalation (PE) and remote code execution (\RCE) exploits. These chains are executable and can span large networks, where numerous potential exploit combinations arise from the large array of network assets, their hardware, software, configurations, and inherent vulnerabilities.
The chains are automatically generated by \name, an AI‑supported framework for the automated discovery of multi‑step PE and \RCE exploit chains in network environments across arbitrary OS environments and segmented networks. 
Through an LLM-based classification approach, \name describes exploits in Planning Domain Description Language (PDDL). PDDL exploit and network descriptions then leverage off-the-shelf AI planners to efficiently find multiple exploit chains.
\name finds 12 unknown chains on an example network with one known planted three-step chain. A red-team exercise validates the executability on one chain using Metasploit modules.
\name is also efficient, finding an exploit chain in 0.01 seconds in an enterprise network with 83 vulnerabilities, 20 hosts, and 6 subnets.
In addition, \name is scalable, it finds an exploit chain in an industrial network with 114 vulnerabilities, 200 hosts, and 6 subnets in 3.16 seconds. It is also comprehensive, finding 13 exploit chains in 26.26 seconds in the industrial network.
Finally, additional results demonstrate \name{’}s flexibility across different exploit sources, its ability to generalize across diverse network types, and its robustness in discovering chains even under constrained privilege assumptions. 
\end{abstract}

%
\IEEEpeerreviewmaketitle

\section{Introduction}\label{sec:intro}
Modern networks are complex, heterogeneous environments with numerous interconnected hosts, each running different operating systems, applications, and services.
While this diversity enhances flexibility and scalability, it also significantly expands the attack surface available to threat actors.
Attackers rarely compromise critical assets through a single vulnerability; many attacks consist of multiple stages, employing interdependent exploits to move laterally through the network, and escalate privileges to expand control~\cite{hutchinsetal}.
For instance, Ransomware and Advanced Persistent Threat (APT) campaigns have plagued modern enterprises, causing significant financial losses~\cite{li2023nodlink}. They often rely on Remote Code Execution (\RCE) and Privilege Escalation (PE) exploits to compromise systems in a sequence of steps~\cite{WONG2023103140}. 
An attacker gains initial access to one host using an \RCE exploit and then escalates privileges using a PE exploit, either on the same host or a different one.
Exploit chaining occurs when the acquired privileges of an exploit are compatible with the required privileges of a subsequent exploit---\RCE exploits can enable PE exploits and, in some cases, vice versa. 
These exploit chains facilitate lateral movement within the network, allowing attackers to compromise additional hosts.
They also enable a technique known as \emph{pivoting}, where an attacker circumvents network segmentation or internal firewalls to move deeper into a network. After gaining initial access to a host on the network, pivoting allows an attacker to access other hosts on the network that would otherwise be unavailable~\cite{kim2023extending}.  

Network hosts can be scanned for vulnerable hardware, operating systems, and applications to reduce the risk of exploits. Scanning, however, can reveal a large number of products requiring patches or updates, and even when vulnerabilities are prioritized, it fails to account for how vulnerabilities might relate to each other through exploit chains~\cite{collins2023identifying} that can lead to the attacker gaining control over a system~\cite{zhang2025scrutinizer}. 
While resources such as the Common Vulnerabilities and Exposures (CVE) database~\cite{nvd-cve}, the Common Vulnerability Scoring System (CVSS), and Common Platform Enumeration (CPE)~\cite{nvd-cpe} of the National Vulnerability Database (NVD) provide insights into individual product vulnerabilities, they describe vulnerabilities in isolation, overlooking the risk posed by multiple vulnerabilities linked together to make exploit chains. Network security teams would benefit from knowing whether their networks are vulnerable to chains of known PE and \RCE exploits.

A scalable and efficient generation of exploit chains that use both PE and \RCE exploits significantly extends prior work that is limited to PE chains on a single host~\cite{299659} or \RCE chains across a network~\cite{obes2013attack}. In addition, unlike traditional vulnerability scanning, such exploit chain generation can reveal complex, multi-step attack paths, enhancing both network risk assessment and the effectiveness of penetration testing.

However, generating viable PE and \RCE exploit chains is a non-trivial task. The possible combinations of known exploits, each with specific prerequisites and effects, result in a combinatorial explosion of possible chains. Discovering these chains requires systematic link-tracing between product configurations (e.g., CPEs), vulnerabilities (e.g., CVEs), whose volume grows yearly~\cite{10179447}, and corresponding exploits. Performing this process manually is time-consuming, error-prone, and does not scale with the increasing size and complexity of modern networks, as well as the rapid growth of the number of vulnerabilities.
To address this challenge, a software-based, semi-automatic method for discovering exploit chains must model the preconditions of each exploit and the privileges gained after it is executed. 
While some of this information can be extracted from textual descriptions of exploits, associated CVEs, or configuration data, other details may have to be inferred by understanding how the exploit works. Additionally, the method must also model relevant details of the network, such as host connectivity and service exposure.


In this paper, we generate executable PE and \RCE chains in networks at scale, both regarding the number of exploits and network size. Our method, \name, can efficiently discover executable PE and \RCE exploit chains in a specific network based on AI planning and supported by a Large Language Model (LLM).
\name starts from a network description, which can be provided by a network security specialist who can then specify a starting host and a target host for potential exploit chains.
\name models the exploit chain discovery problem as an AI planning problem, beginning with the classification of known exploits from an exploit source library, including their prerequisites and the resulting privileges they confer. The classification can be provided manually or aided by LLMs.
To use the AI-planning approach, \name models the exploits as actions and network elements as objects. 
After specifying a goal, from an initial host to a target host, \name uses an off-the-shelf AI planner to generate plans that are essentially executable exploit chains.

Our contributions are as follows.
\begin{enumerate}
\item \name is the first method capable of discovering exploit chains of existing PE and \RCE exploits on segmented networks.
\item We demonstrate the effectiveness, efficiency, scalability, and robustness of generating exploit chains with \name by using sample networks with planted vulnerabilities. As evidence of
\begin{itemize}
    \item \emph{effectiveness}, \name is able to find \textbf{multiple} chains in the same network. In a network of 200 hosts and 6 subnets, which has 114 vulnerabilities, 13 exploit chains can be found in 26.25 seconds.
    \item \emph{efficiency}, \name is able to find an exploit chain in a network of 20 hosts and 6 subnets, which has 83 vulnerabilities in 0.01 seconds.
    \item \emph{scalability}, \name is able to find an exploit chain in a network of 200 hosts and 6 subnets, which has 114 vulnerabilities in 3.16 seconds.
    \item \emph{robustness}, \name is able to discover exploit chains even when assuming that services are configured with the lowest privileges possible.
    \end{itemize}
\item We demonstrate that \name achieves broad and flexible exploit chain discovery, supporting both on‑server and cross‑network analysis, and accommodating heterogeneous sources of exploits. 
\begin{itemize}
    \item \name can discover unknown exploit chains across hosts and subnets, achieving both on-server and cross-network chain detection. We set up a sample network with vulnerabilities and a single known chain. In addition to the known chain, \name discovers 12 more chains on this network that were unknown, and we demonstrate the manual execution of one of them. 
    \item \name is general and able to model exploits from multiple sources, i.e., within a framework or a library. We demonstrate our implementation of \name on Metasploit~\cite{metasploit} with 1,880 exploits and the \coreblah~\cite{coresec} with 1,903 exploits.
\end{itemize}

\end{enumerate}

We proceed as follows. Section~\ref{sec:background} provides background. Section~\ref{sec:motivation} motivates \name with an example and describes its threat model. Section~\ref{sec:method} provides a technical description of \name.
Section~\ref{sec:demonstration} validates \name using the motivating example. Section~\ref{sec:evaluation} analyzes \name through various experiments. Section~\ref{sec:discussion} presents the discussion and future work.
Section~\ref{sec:related_work} discusses related work and similar systems.
Finally, Section~\ref{sec:conclusion} concludes and summarizes the results of this paper.

\section{Background}\label{sec:background}
 We elaborate upon exploit chains in Section~\ref{sec:background:chains} and provide an overview of AI planning as it relates to \name in Section~\ref{sec:background:planning}.

\subsection{Exploit Chains}
\label{sec:background:chains}
An \textbf{exploit chain} is a multi-step sequence intended to compromise a target host~\cite{299659}. It entails the sequential execution of two or more interdependent exploits. 
The result of each exploit, combined with system vulnerabilities, enables the attacker to execute subsequent exploits to advance progressively toward their goal.
In this paper, we focus on exploit chains connecting \textbf{Remote Code Execution (\RCE) exploits} and \textbf{Privilege Escalation (PE) exploits} that enable an attacker to gain access, move laterally, execute commands, and/or escalate privileges as defined by MITRE ATT\&CK Matrix~\cite{mitre}.

\emph{\RCE exploits} target vulnerabilities in web applications or network services that use communication protocols such as TCP and UDP, enabling an attacker to run code on a remote machine~\cite{285501}. This type of exploit can result in gaining control of the affected system with the same (or higher) privileges as the exploited process, potentially establishing a foothold for subsequent exploits.
In contrast, \emph{PE exploits} target vulnerabilities in operating systems, kernels, and misconfigured or buggy software applications~\cite{287202}. This type of exploit enables the attacker to gain control of the host where they have established a foothold, elevating their privileges to access sensitive data or disrupt services.

\subsection{AI Planning}
\label{sec:background:planning}
\textbf{AI Planning} is a method for automatically deriving a sequence of operative actions to reach a goal. An AI Planner, starting from the current state, enumerates an optimal or feasible path to a goal state through a set of possible actions, considering constraints and available resources. 

The \textbf{Planning Domain Definition Language} (PDDL)~\cite{299659} is a language used to define planning tasks. Each planning task is modeled through a PDDL domain file and a PDDL problem file. While the first specifies the object types, predicates, and possible actions in a goal-directed planning task, the latter lists the existing objects, initial conditions, and the desired goal of the associated task. 
Together, the domain and problem provide a complete specification of a planning task. 
Each action in a PDDL domain file includes parameters, preconditions, and effects. The parameters define the object types that participate in the action. Preconditions specify the logical conditions, expressed as predicates, that must hold true for the action to be executed. Effects describe the logical changes that occur as a result of the action, capturing the new state of the system. 
Predicates, on the other hand, represent conditions or relationships that can be true or false. They accept arguments and can be combined into complex Boolean formulas to articulate the preconditions and effects of actions. 

In the so-called \emph{classical} planning tasks, actions can modify the truth values of predicates both positively and negatively, enabling or disabling conditions as needed. In contrast, exploit chain discovery aligns more naturally with \emph{monotonic} planning, wherein predicates can only transition to the true state.
This reflects the typical behavior in security contexts, where once privileges are gained, they are rarely revoked during the attack sequence. Formally, a monotonic planning task can be represented by a tuple \((S, A, \gamma, s_0, s_G)\):
\begin{itemize}
\item \(S\): the set of all the possible states of the predicates.
\item \(s_0 \subset S\): the initial state.
\item \(s_G \subseteq S\): the goal state.
\item \(A\): the set of actions \(a\), where each action \(a \in A\) has:
\begin{itemize}
  \item Preconditions: $\textnormal{pre}(a) \subset S$, conditions required for \(a\) to be applied.
  \item Effects: $\textnormal{eff}(a) \subset S$, effects of the action \(a\), indicating the conditions that become true after $a$ is applied.
\end{itemize}
\item \(\gamma(s, a) \rightarrow s'\): the transition function defining state \(s'\) reached from state \(s\) when action \(a\) is applied, such that \(s'=s\cup \textnormal{eff}(a)\), provided that \(\textnormal{pre}(a)\subseteq s\)
\end{itemize}

The objective is to find a sequence of actions (a plan) \(\pi = (a_1, a_2, \dots, a_n)\), where each \(a_i \in A\), starting from \(s_0\), applying \(\pi\) leads to a state \(s_G \subseteq S\).

This framework enables precise and logical descriptions of system dynamics, allowing AI planners to find action sequences to achieve specific goals. 
In \name, each exploit is modeled as a planning action, where the action's preconditions represent the requirements for the execution of the exploit, such as the attacker’s current privileges and the presence of a vulnerable configuration on the target host. 
The action’s effects specify the system state resulting from the exploit, reflecting the privileges acquired after executing the exploit. The plan generated by the AI planner constitutes an exploit chain that achieves the attacker’s objective.


\section{Motivation and Threat Model}\label{sec:motivation}
In Section~\ref{sec:motivation_threatmotivation_threat:threat} the threat model is discussed, detailing the assumptions and scope of our approach. In Section~\ref{sec:motivation_threat:example}, we present an example that motivates \name.

\subsection{Threat Model}
\label{sec:motivation_threatmotivation_threat:threat}

The objective of \name is to identify feasible exploit chains within a known network configuration from a defender's perspective. 

Our threat model operates under the following key assumptions:
\begin{itemize}
    \item \textbf{Network Knowledge}: the topology, host configurations (hardware, operating systems, and software), and interconnections are known. This reflects a defender with complete visibility of the infrastructure.
    \item \textbf{Static Environment}: the network state and host configurations remain static during planning. Basic firewall configurations (e.g., blocked traffic to specific ports) are supported, but are assumed to remain unchanged throughout the analysis. 
    This assumption does not hinder the applicability of \name, since it is designed to perform exploit chain discovery fast enough that changes to the network topology or configurations are unlikely to occur within the analysis timeframe.
    \item \textbf{Scope}: we focus on two exploit \textbf{types} (\RCE and PE), two \textbf{transport protocols} (TCP and UDP), and four \textbf{privileges} levels (None, Low, High, and Root) described in Table~\ref{tab:privilege}.
    \item \textbf{Monotonic planning problem}: we model the exploit chain discovery process as a monotonic planning problem, where the attacker can only gain privileges on any host. While this abstraction simplifies real-world privilege systems, it provides a practical framework for planning and analyzing exploit chains.
\end{itemize}

\begin{table}[tb]
    \centering
    \caption{Privilege Levels. The privilege model used by \name. An attacker is assumed to have one of these privileges on each host in the network at each step of the exploit chain.}
    \begin{tabular}{p{0.25\linewidth}|p{0.65\linewidth}}
    \textbf{Privilege Level} & \textbf{Description}\\ \hline
    None (N) & The attacker has no access to the host. They may interact with it over the network.\\\hline
    Low (L) & The attacker has unprivileged user-level control on the host. They can execute basic commands or run non-sensitive applications with limited permissions, but likely cannot access many files or run many executables. \\ \hline
    High (H) & The attacker has privileges equivalent to an administrator on the host. They can run arbitrary code, but they still lack full system-level control. \\ \hline
    Root (R) & The attacker has full system control on the host (e.g., root on Unix-like systems or NT Authority\textbackslash SYSTEM on Windows). \\
    \end{tabular}
    \label{tab:privilege}
\end{table}

The attacker aims to gain privileged command execution (e.g., root privilege level) on a designated target host. To do so, they use known (i.e., explicitly available and documented) exploits to gain control and escalate privileges across hosts in the network. Each exploit is modeled as an atomic planning action in the PDDL domain file with its specific preconditions (e.g., low privilege level on the host, exposed service) and corresponding effects (e.g., increased privileges or access to a new host).
The attacker operates within the perimeter defined by the PDDL problem file, which may describe a partial or complete view of the network depending on the set of hosts under analysis. This flexibility allows \name to model different attacker scenarios, ranging from external adversaries targeting the entire network to internal threats acting within a specific subnet. In our evaluation, we always consider the case of an external attacker.

\name is designed to generalize across various environments, including traditional enterprise networks, containerized clusters, virtualized platforms, and cloud deployments. 
While our evaluation uses publicly available exploits from Metasploit~\cite{metasploit} to validate the discovered exploit chains, the framework is not tied to a specific source. \name can be extended to use other exploit repositories (e.g., proprietary or private), provided that sufficient information for the classification of the exploits is available. The use of Metasploit enables validation of exploit chains on testbeds, ensuring their practical feasibility.

\subsection{Motivating Example}
\label{sec:motivation_threat:example}
\begin{figure}[ht]
    \centering
    \includegraphics[scale=0.95]{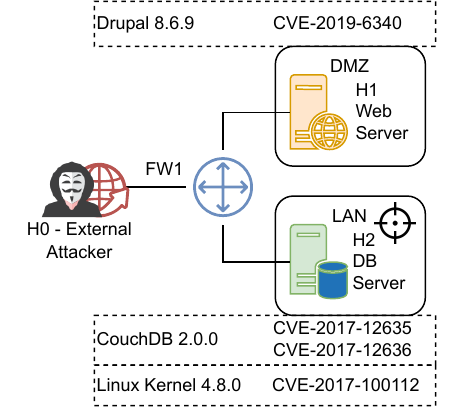}
    \caption{Motivating example. A sample network with DMZ and LAN architecture and a threat actor on the internet on H0. \hostone runs a vulnerable web server. \hosttwo runs a vulnerable database server. \hostone, is affected by CVE-2019-6340 (Drupal 8.6.9). \hosttwo is affected by CVE-2017-12635 and CVE-2017-12636 (CouchDB 2.0.0), along with CVE-2017-1000112 (Linux Kernel). An exploit chain of length three allows a threat actor to gain root command execution on \hosttwo.}
    \label{fig:motivating_example_network}
\end{figure}

To demonstrate how attackers can exploit chained vulnerabilities across segmented networks, we present the motivating example featuring a DMZ–LAN architecture shown in Figure~\ref{fig:motivating_example_network}. This setup is divided into two subnets:

\begin{itemize}
    \item Demilitarized Zone (DMZ): controls the exchange of data between the Internet and the internal network, hosting public-facing services such as web servers.
    \item Local Area Network (LAN): hosts internal services and sensitive assets not directly exposed to the Internet, such as a database server.
\end{itemize}

The DMZ contains one host (\hostone) configured as a web server running \textbf{Ubuntu Linux 16.04}, \textbf{Drupal 8.6.9}, and \textbf{PHP 7.0.33}
The LAN contains one host (\hosttwo) configured as a database server running \textbf{Linux Kernel 4.8.0} and \textbf{Apache CouchDB 2.0.0}.
\hostzero represents the external attacker, whose objective is to gain root command execution on \hosttwo.
This configuration reflects a realistic, vulnerable web application architecture composed of a web server front-end and a database back-end. 

\begin{figure*}[h!]
    \centering
    \includegraphics[width=0.95\linewidth]{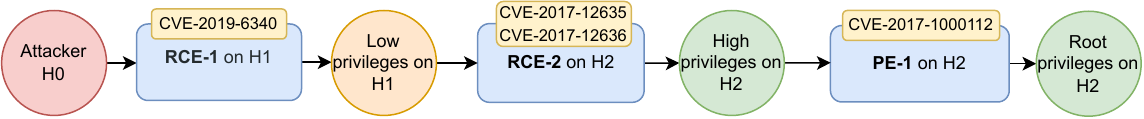}
    \caption{Motivating example exploit chain in a network with DMZ and LAN architecture. Rounded blue boxes represent an exploit, and circles denote privileges acquired after an exploit (the effect of an action). Yellow boxes represent the vulnerabilities that allow the related exploit to be executed (the preconditions of an action).}
    \label{fig:motivating_example_exploit_path}
\end{figure*}

Those vulnerabilities enable the execution of the exploit chain illustrated in Figure~\ref{fig:motivating_example_exploit_path}. 
The attacker first executes the \texttt{Drupal RESTful Web Services unserialize() RCE}~\cite{RCE_DRUPAL} exploit (\RCE-1) on \hostone in the DMZ to gain user-level access, exploiting CVE-2019-6340~\cite{CVE-2019-6340}.
Next, using the \texttt{Apache CouchDB Arbitrary Command Execution} exploit~\cite{RCE_COUCHDB} (\RCE-2), the attacker moves laterally into the LAN and obtains high privileges on \hosttwo, exploiting CVE-2017-12635~\cite{CVE-2017-12635} and CVE-2017-12636~\cite{CVE-2017-12636}.
Finally, the attacker deploys the \texttt{Linux Kernel UDP Fragmentation Offset (UFO) Privilege Escalation} exploit~\cite{PE_LinuxKernel} (PE-1) on \hosttwo to escalate privileges to root, exploiting CVE-2017-1000112\cite{CVE-2017-1000112}.

\section{Technical Description}\label{sec:method}
\begin{figure*}[tb]
    \centering
    \includegraphics[width=0.99\linewidth]{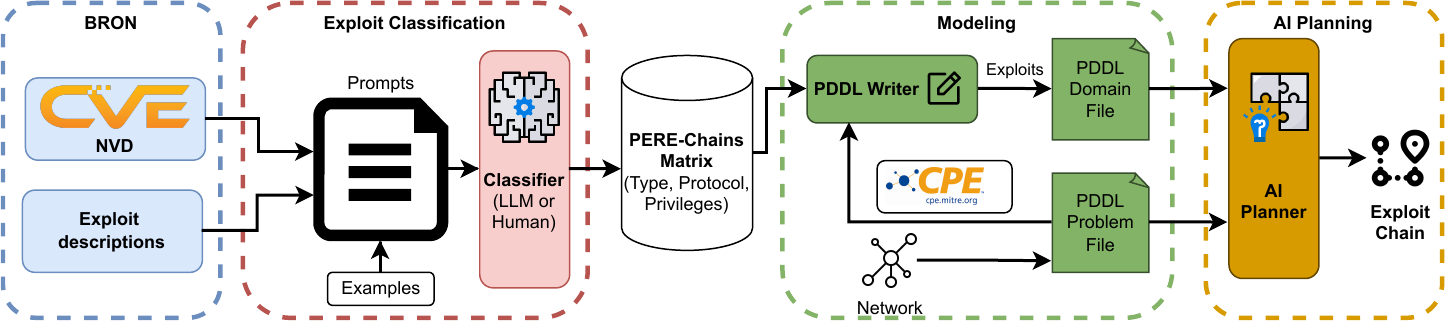}
    \caption{\textbf{System architecture of \name}. \name is divided into three major steps: (i) Exploit classification, (ii) Modeling, and (iii) AI Planning. In step (i), the method processes information from the NVD and exploit descriptions using a classifier (human or LLM-assisted) to extract the exploit type, transport protocol (for \RCE exploits), required privileges, and acquired privileges. This information is stored in the \name Matrix. In step (ii), a human user models the network in a PDDL Problem File. A PDDL writer then selects the relevant exploits and encodes them in a PDDL domain file. In step (iii), an off-the-shelf AI planner solves the planning problem using the PDDL problem and domain Files and outputs feasible PE and \RCE exploit chains across the network.}
    \label{fig:overview}
\end{figure*}

Figure~\ref{fig:overview} shows a overview of \name framework, which has three main stages:
\begin{itemize}
    \item[\textbf{1 - \acmatrix{.}}] The first step in \name is to classify known exploits by type (PE or \RCE exploits), communication protocol (for \RCE exploits), required execution privileges, and privileges acquired upon successful execution.
    We classify both structured and unstructured text data extracted from publicly available exploit and vulnerability sources (e.g., Metasploit, NVD). 
    The results of the classification are stored in the \name matrix, which provides a structured foundation for subsequent PDDL modeling.
    \item [\textbf{2 - \acmodel{.}}] To enable automated planning, \name models the target environment and the exploit set in PDDL. The network topology, host configurations, and attacker objectives are manually encoded into a PDDL problem file; while this process could be automated, it is not central to the proposed approach. Exploits stored in the \name matrix are programmatically translated into PDDL domain actions, where preconditions capture host-specific requirements (e.g., software product, network protocol, and attacker privileges), and effects reflect the privilege escalation or remote access granted by the exploit. These actions are compiled into a PDDL domain file.
    \item [\textbf{3 - \acplan{.}}] To obtain exploit chains, we run an AI planner on the PDDL domain and problem files, where we can define initial and goal states. The resulting plans correspond to exploit chains. 
\end{itemize}

The advantage of using public exploit module libraries, such as the one in Metasploit, is that discovered exploit chains can be automatically validated using the framework.

\subsection{Exploit Classification}
\label{sec:method:classification}

\name uses exploit and vulnerability information to classify exploits. Each exploit in \name is classified by assigning it a type (PE or \RCE), a transport protocol (TCP or UDP; only for \RCE exploits), the privileges required to run it (None, Low, or High), and the privileges acquired by running it (Low, High, or Root). This results in one of fifteen possible exploit classes, which are enumerated in Table~\ref{tab:exploit-classifications}. The privilege levels used in our model are described in Table~\ref{tab:privilege}.

\begin{table}[h]
    \centering
    \caption[Exploit Classifications]{Exploit Classifications. In the \name model, all exploits can be classified into one of the following 15 classes. We have abbreviated the privilege level names None (N), Low (L), High (H), and Root (R). The first privilege corresponds to the privileges required on the target host to run the exploit. The second privilege corresponds to the privileges gained on the target host after running the exploit.}
    \begin{tabular}{lll}
    \multicolumn{3}{c}{\textbf{Exploit Classes}}                                               \\ \hline
    \multicolumn{1}{l|}{PE\_L\_H}       & \multicolumn{1}{l|}{PE\_L\_R}       & PE\_H\_R       \\
    \multicolumn{1}{l|}{RCE\_TCP\_N\_L} & \multicolumn{1}{l|}{RCE\_TCP\_N\_H} & RCE\_TCP\_N\_R \\
    \multicolumn{1}{l|}{RCE\_TCP\_L\_H} & \multicolumn{1}{l|}{RCE\_TCP\_L\_R} & RCE\_TCP\_H\_R \\
    \multicolumn{1}{l|}{RCE\_UDP\_N\_L} & \multicolumn{1}{l|}{RCE\_UDP\_N\_H} & RCE\_UDP\_N\_R \\
    \multicolumn{1}{l|}{RCE\_UDP\_L\_H} & \multicolumn{1}{l|}{RCE\_UDP\_L\_R} & RCE\_UDP\_H\_R
    \end{tabular}
    \label{tab:exploit-classifications}
\end{table}

\subsubsection{Exploit and Vulnerability Data Sources}

In our implementation of \balfachains, we use \textbf{BRON}~\cite{BRON} to collect information about exploits and vulnerabilities. BRON is a property graph database that integrates different cybersecurity databases~\cite{exploitdb}. 
It provides access to vulnerabilities as described by the NVD CVE data.
Every CVE entry describes a vulnerability with a CVE identifier, semi-structured text, and lists the software configurations known to have the vulnerability.
These configurations are specified using the CPE, a naming scheme for hardware and software products. The latest version of CPE, version 2.3, provides information about part (application, operative system, or hardware), vendor, product, version, update, edition, language, software edition, target software, target hardware, and others~\cite{mitre-cpe}, e.g.,   cpe:2.3:a:apache:couchdb:2.0.0:*:*:*:*:*:*:* represents the CouchDB application, developed by Apache, version 2.0.0. The asterisks represent the logical value ANY for the update, edition, language, software edition, target software, target hardware, and others.

For this study, we use 1,880 exploits available in Metasploit, which can be easily and quickly executed in exploit chains using the Metasploit Framework, along with 2,002 CVEs linked by BRON.
We also use 1,903 exploits from the \coreblah, a component of the Core Impact penetration testing platform, where tested and validated exploits are collected and maintained by Fortra’s Core Security team\footnote{\url{https://www.fortra.com/}}.

\subsubsection{Classification}
\label{sec:class-meth}
Classifying exploits is the initial step in \balfachains. It can be performed manually or with the assistance of a LLM---GPT-4o in our case. For each exploit, the LLM processes the exploit's text description, the associated vulnerabilities' text descriptions, and the vulnerable configurations specified through CPEs.

Effective prompt engineering is crucial for obtaining accurate and reliable LLM results. We employ well-structured and precise prompts that provide a clear context for the model to follow and stay on topic with responses~\cite{chen2023unleashing}. In our approach, we utilize a system prompt that explicitly defines privilege levels and provides detailed descriptions of each exploit class. To enhance the classification process, we employ few-shot prompting~\cite{brown_language_2020} with six carefully selected examples. This technique leverages the model’s in-context learning capability~\cite{dong-etal-2024-survey}, enabling it to adapt to new tasks efficiently by learning from examples embedded directly within the prompt.

The prompt provided to the LLM for classification includes the following components in order:
\begin{itemize}
    \item \textbf{Exploit text}: textual description that comes with each exploit (e.g., descriptions from Metasploit or \coreblah).
    \item \textbf{CVE texts}: textual descriptions of the vulnerabilities related to an exploit (e.g., descriptions from NVD).
    \item \textbf{Configurations}: list of configurations vulnerable to the exploit (i.e., CPEs).
\end{itemize}

LLM-inferred classifications of required privilege are improved by using extractions from the CVSS 3.x~\cite{nvd-cvss} vectors, which provide details about the privileges required for an exploit (None, Low, High). However, earlier versions of CVSS lack this information, and certain critical classification elements, such as privileges acquired and protocol, are not explicitly defined in CVSS.

\subsection{Modeling}
\label{sec:exploit-modelling}
Exploit modeling in \name consists of generating the PDDL problem and domain file, which represent the network environment and the set of applicable exploits, respectively. 
The \textbf{PDDL problem file} describes the network topology, the configuration of each host (expressed using CPE-like attributes), and specifies the initial and target hosts.
Network connectivity is modeled with the predicate \texttt{(connected\_to\_network ?h - Host ?n - Network)}, while host configurations are represented using predicates that map to CPE attributes.
Notably, version information is broken into semantic components using \texttt{(has\_version ?h - Host ?p - Product ?ma - Major ?mi - Minor ?pa - Patch)}, distinguishing between major, minor, and patch. 
The attacker’s initial foothold is specified using the predicate \texttt{(is\_compromised attacker\_host agent ROOT\_PRIVILEGES)}, and the attack objective is defined via a goal clause such as \texttt{(:goal (is\_compromised target\_host agent ROOT\_PRIVILEGES))}. Since this information is under the control of the network security specialist, we assume it is manually prepared; however, it could be automated in the future through a more convenient interface or API.

The \textbf{PDDL domain file} relevant to the problem is automatically generated by \name{'}s \textbf{PDDL Writer}, which converts entries from the \name matrix into planning actions. 
Each action corresponds to a real-world exploit and encodes its logical preconditions (e.g., required privileges, vulnerable configuration, connectivity) and effects. 
To optimize performance, the PDDL Writer includes only the exploits for which the hosts in the network under study are vulnerable. Additionally, it prunes configuration-related preconditions to match only those present in the specific problem instance, thereby reducing complexity and improving efficiency in the subsequent AI Planning step.

\name uses a structured set of types and predicates to describe the planning task.
Table~\ref{tab:pddl-types} presents the object types used in the domain, categorized into general entities (e.g., hosts, agents), configuration descriptors (e.g., products, editions), and versioning.
Table~\ref{tab:pddl-predicates} lists the predicates that define the state of a host (e.g., \texttt{is\_compromised}), its connectivity (e.g., exposed services and connections to subnets), and its configuration (e.g., CPEs' information).
This design is inspired by the work in~\cite{obes2013attack}, where they use a similar model. Contrary to ~\cite{obes2013attack}, we avoid modeling ports directly and instead specify whether a service is listening on any port through the predicates \verb |TCP_listen| and \verb |UDP_listen|. Additionally, our \verb|TCP_connected| and \verb|UDP_connected| predicates are service-specific. This allows us to model cases where a host can send traffic to a remote host, but only to specific services. 

\begin{table}[tb]
    \centering
    \caption{PDDL Types. These are the types used when writing PDDL.}
    \begin{tabular}{p{0.3\linewidth}|p{0.6\linewidth}}
    \textbf{Category} & \textbf{Types} \\\hline
    General Model & Network, Agent, Host, Privilege  \\\hline
    Configuration & 
    Product, Version, Update, Edition, Language, SW\_Edition, Target\_SW, Target\_HW, Other \\\hline
    Versioning & Major, Minor, Patch\\
    \end{tabular}
    \label{tab:pddl-types}
\end{table}

\begin{table}[tb]
\centering
\caption{PDDL Predicates. These are the predicates used when writing PDDL.}
\begin{tabular}{l|ll}
\multicolumn{1}{c|}{\textbf{Category}} & \multicolumn{2}{c}{\textbf{Predicates}} \\ \hline
Compromise & is\_compromised & \\ \hline
\multirow{3}{*}{Connection} & TCP\_listen & UDP\_listen \\
 & TCP\_connected & UDP\_connected \\
 & connected\_to\_network \\
 & trusted\_channel & \\\hline
\multirow{5}{*}{Configuration} & has\_product & has\_version \\
 & has\_update & has\_edition \\
 & has\_language & has\_sw\_edition \\
 & has\_target\_sw & has\_target\_hw \\
 & has\_other & \\
\end{tabular}
\label{tab:pddl-predicates}
\end{table}

\subsection{AI Planning}
\label{sec:method:planning}
\name passes the generated PDDL problem and domain files to an AI planner, which returns sequences of actions representing exploit chains. 
This step is planner-agnostic and supported by multiple off-the-shelf planners. In our evaluation, we test \name using the Fast Downward planning system~\cite{helmert2006fast} and ENHSP~\cite{scala2016interval}, two widely used frameworks in the AI planning community.
When using Fast Downward, we evaluate three different configurations: LAMA~\cite{richter2010lama}, LAMA-first, and K*~\cite{lee-et-al-socs2023}.

The K* planner is capable of returning multiple valid plans for a single planning task, compared to the others, which can return only a single plan. This feature is essential in the context of \name, as it enables the discovery of multiple distinct exploit chains that achieve the same security objective, such as compromising a specific target host, given a starting point.

\section{Validation of \name on Motivating Example}\label{sec:demonstration}
In this section, we validate \name by running it on the motivating example from Section~\ref{sec:motivation_threat:example}, which contains three planted vulnerabilities that can be chained together to gain ROOT\_PRIVILEGES over the database server (\hosttwo) from an attacker host placed outside of the network (\hostzero).

\paragraph{\textbf{Exploit Classification}} BRON~\cite{hemberg2021using} is used to extract the text descriptions from Metasploit, the related CVEs, and the vulnerable configurations in the CPE form. In addition to the three planted vulnerabilities, BRON returned 17 other relevant exploits that could affect the hosts due to their configurations and be involved in potential exploit chains, for a total of 20 exploits relevant to the problem.
In the exploit classification step, \name identified the appropriate type of exploit, protocol, required privileges, and acquired privileges for the planted chain, as shown in Table~\ref{tab:mot_ex_class}. 
\begin{table}
    \centering
    \caption{Exploit type, protocol, required, and acquired privileges classification selected by \name for the three exploits in the motivating example~(Fig~\ref{fig:motivating_example_network})}
    \begin{tabular}{c|c|c}
        \textbf{Exploit} & \textbf{Classification} & \textbf{Ground Truth}\\ \hline
        \RCE-1 & RCE\_TCP\_N\_L & RCE\_TCP\_N\_L \\
        \RCE-2 & RCE\_TCP\_N\_H & RCE\_TCP\_N\_H \\
        PE-1 & PE\_L\_R & PE\_L\_R \\
    \end{tabular}
    \label{tab:mot_ex_class}
\end{table}

\paragraph{\textbf{Modeling}} In the \name exploit modeling step, the problem and the relevant exploits are described in PDDL.
An excerpt from the PDDL problem file is shown in Listing~\ref{pddl_problem_list} to show the initial state, the network topology, the configuration of \hostone and \hosttwo with the specification of the products, versions, and services exposed, and finally the goal state. 

\begin{lstlisting}[label=pddl_problem_list,caption=PDDL description of hosts in the motivating example network,float,frame=tb, backgroundcolor = \color{lightlightgray},captionpos=b]
;; INITIAL STATE
(is_compromised attacker_host agent ROOT_PRIVILEGES)

;; NETWORK TOPOLOGY
(connected_to_network attacker_host dmz)
(connected_to_network web_server dmz)
(connected_to_network web_server lan)
(connected_to_network db_server lan)

;; HOST 1
(has_product web_server o--canonical--ubuntu_linux)
(has_version web_server o--canonical--ubuntu_linux ma16 mi4 pa0)
(has_product web_server a--drupal--drupal)
(has_version web_server a--drupal--drupal ma8 mi6 pa9)
(TCP_listen web_server a--drupal--drupal)
(has_product web_server a--php--php)
(has_version web_server a--php--php ma7 mi0 pa33)

;; HOST 2
(has_product db_server o--linux--linux_kernel)
(has_version db_server o--linux--linux_kernel ma4 mi8 pa0)
(has_product db_server a--apache--couchdb)
(has_version db_server a--apache--couchdb ma2 mi0 pa0)
(TCP_listen db_server a--apache--couchdb)

;; GOAL STATE
(:goal (is_compromised db_server agent ROOT_PRIVILEGES))
\end{lstlisting}

Listing~\ref{couchdb_pddl} shows the PDDL action that models the exploit \RCE{-2}. Each PDDL action requires a name, a set of parameters, a list of logically connected preconditions, and a defined effect.
Since this is an \RCE exploit, the action includes two parameters of type Host: \texttt{?local\_host}, representing the origin of the attack, and \texttt{?remote\_host}, the target.
The preconditions can be grouped into two main categories. The first specifies the privilege level required on \texttt{?local\_host}, in this case, any privilege level higher than None.
The second defines the network reachability and service exposure: \texttt{?local\_host} must be able to reach \texttt{?remote\_host} over TCP via \texttt{a--apache--couchdb}, and the target must be running a vulnerable version of the product (e.g., \texttt{ma2 mi0 pa0}, corresponding to version 2.0.0).
Finally, the effect of the action models the outcome of a successful exploit execution: the agent \texttt{?agent} gains \texttt{HIGH\_PRIVILEGES} on \texttt{?remote\_host}.

\begin{lstlisting}[label=couchdb_pddl,caption=\RCE-2 PDDL action,float,frame=tb, backgroundcolor = \color{lightlightgray},captionpos=b]
(:action couchdb_rce
  :parameters (?local_host - host 
               ?remote_host - host 
               ?agent - agent)
  :precondition (and
   (or
    (is_compromised ?local_host ?agent LOW_PRIVILEGES)
    (is_compromised ?local_host ?agent HIGH_PRIVILEGES)
    (is_compromised ?local_host ?agent ROOT_PRIVILEGES)
   )
   (TCP_connected ?local_host ?remote_host a--apache--couchdb)
   (has_product ?remote_host a--apache--couchdb)
    (has_version ?remote_host a--apache--couchdb ma2 mi0 pa0)
  )
  :effect (is_compromised ?remote_host ?agent HIGH_PRIVILEGES)
 )
\end{lstlisting}

\paragraph{\textbf{AI Planning}} In the \name AI planning step, the PDDL problem and domain files were provided to the Fast Downward planner using the K* heuristic configuration. The planner successfully generated 13 plans. Listing~\ref{lst_me_chain_orig} shows the specific exploit chain we expected to find based on the planted vulnerabilities in the network.

\begin{lstlisting}[label=lst_me_chain_orig,caption=The Exploit chain plan for the motivating example,float,frame=tb,backgroundcolor = \color{lightlightgray},captionpos=b]
tcp_connect dmz attacker_host web_server a--drupal--drupal agent (1)
drupal_restful_web_service attacker_host web_server a--drupal--drupal agent (1)
tcp_connect lan web_server db_server a--apache--couchdb agent (1)
apache_couchdb_arbitrary_command_execution web_server db_server a--apache--couchdb agent (1)
linux_kernel_udp_fragmentation_offset_ufo_pe db_server agent (1)
\end{lstlisting}

We then deployed the motivating example in a controlled test environment using VirtualBox.
To validate the plan, we manually executed each exploit in the chain using Metasploit modules, invoking them sequentially in the terminal with the appropriate options and payloads.
This process involved gaining low privileges on the web server, moving to the database server, and finally escalating to root privileges on the target.

\begin{figure}
    \centering
    \includegraphics[width=0.99\linewidth]{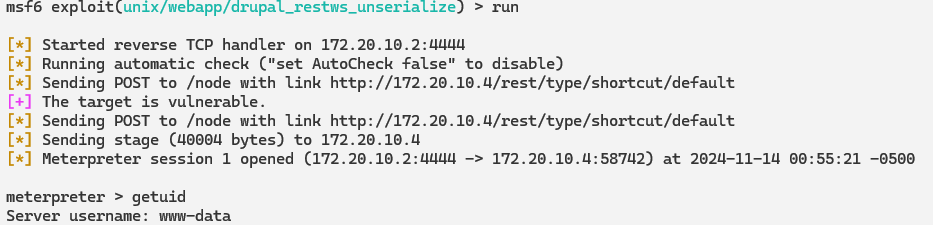}
    \includegraphics[width=0.99\linewidth]{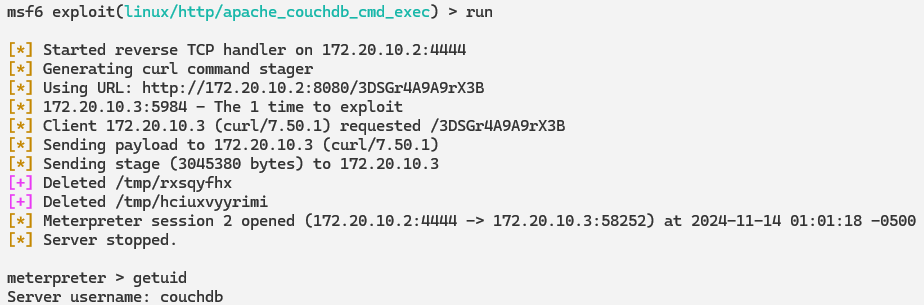}
    \includegraphics[width=0.99\linewidth]{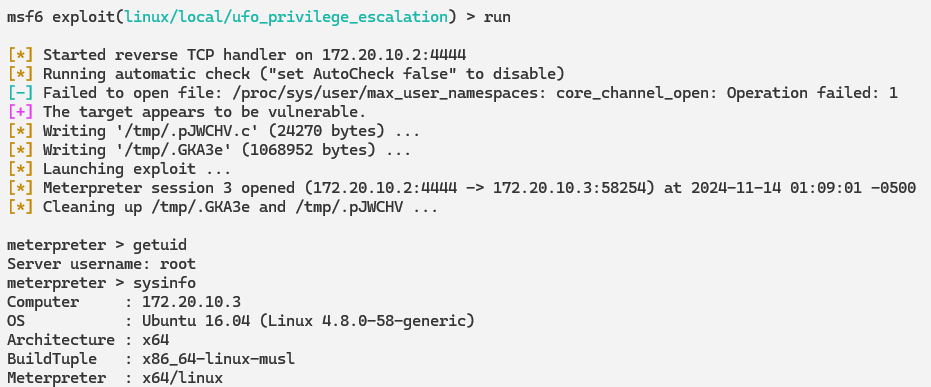}
    \caption{Privileges acquired for each step of the exploit chain in Listing~\ref{lst_me_chain_orig}}
    \label{fig:chain_deploy}
\end{figure}

Figure~\ref{fig:chain_deploy} shows the progression of privilege levels acquired across the virtual machines as each step of the exploit chain was executed.

\begin{figure}
    \centering
    \includegraphics[width=0.99\linewidth]{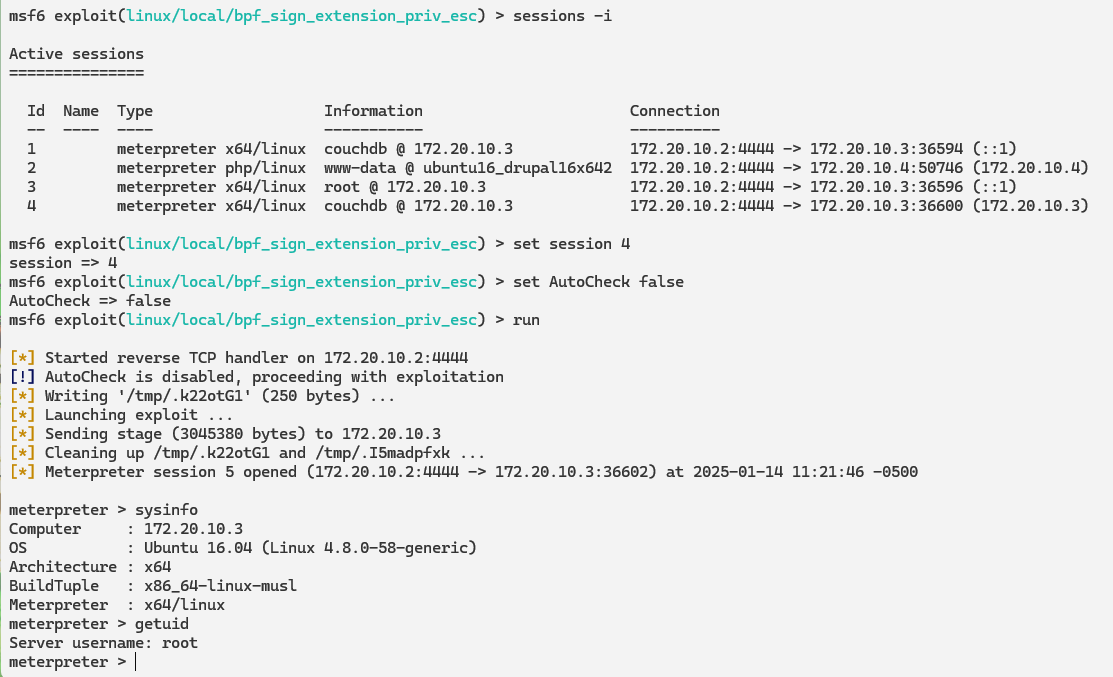}
    \caption{Privileges acquired for the last step of a discovered additional exploit chain, see Listing~\ref{lst_me_chain}}
    \label{fig:chain_deploy_new}
\end{figure}

\name successfully identified the anticipated exploit chain, beginning from \hostzero, where the threat actor has root privileges, and extending to \hosttwo, where root privileges are gained. 
In addition to the expected path, \name discovered 12 alternative exploit chains based on the same network configuration. For instance, one alternative chain follows the same initial steps but concludes with a different privilege escalation exploit, as shown in Listing~\ref{lst_me_chain}. Figure~\ref{fig:chain_deploy_new} illustrates the final step, where the \texttt{bpf\_sign\_extension\_priv\_esc}~\cite{CVE-2017-16995} exploit is executed.

\begin{lstlisting}[label=lst_me_chain,caption=Another Exploit chain plan discovered in the motivating example,float,frame=tb,backgroundcolor = \color{lightlightgray},captionpos=b]
tcp_connect dmz attacker_host web_server a--drupal--drupal agent (1)
drupal_restws_unserialize attacker_host web_server agent (1)
tcp_connect lan web_server db_server a--apache--couchdb agent (1) 
apache_couchdb_cmd_exec web_server db_server agent (1)
bpf_sign_extension_priv_esc db_server agent (1)
\end{lstlisting}

\section{\name Capability Analysis}\label{sec:evaluation}
In this section, we analyze \name{'} capabilities experimentally. We describe our evaluation hardware, network setups, and planners in Section~\ref{sec:evaluation:experimental_setup}. 
In Section~\ref{sec:evaluation:planning} we analyze the AI planning module, while in Section~\ref{sec:evaluation:class} we analyze the exploit classification component. 

\subsection{Experimental Setup}
\label{sec:evaluation:experimental_setup}
We conducted our experiments on commodity hardware running Ubuntu Linux 22.04.1, with an Intel(R) Core(TM) i7-8700K CPU 6 cores @ 3.70GHz chip, and 64 GB of RAM DDR4 @ 2667 MHz. We set up a Docker container running workers with 20GB of RAM to host \texttt{planutils}~\cite{muise-et-al-icaps2022systemdemos}, a general library for setting up Linux-based environments for running planners. To interface with the LLM-based classifier, we used the \texttt{LangChain} library\footnote{\url{https://github.com/langchain-ai/langchain}}, which facilitates structured interactions with language models. To validate the PDDL actions and generated files, we employed the \texttt{pddl} library\footnote{\url{https://github.com/AI-Planning/pddl}}, which provides parsing and consistency-checking capabilities for PDDL domain and problem definitions.

The planners used for the experiments are:
\begin{itemize}
    \item Fast Downward (FD)~\cite{helmert2006fast,HELMERT2009503}: planning system that uses a heuristic search-based approach for solving classical planning problems. 
    We combined FD with the following planners:
    \begin{itemize}
        \item Landmark-Aware Multi-Heuristic Planner (LAMA)~\cite{richter2010lama}: anytime planner that uses a heuristic derived from landmarks in conjunction with the FF heuristic~\cite{hoffmann2001ff}. It first produces a suboptimal solution quickly and then iteratively improves it.
        \item LAMA-first: simplified configuration of LAMA that only focuses on finding the first solution rather than optimizing or improving the plan.
        \item K*~\cite{lee-et-al-socs2023}: planner designed specifically to find multiple optimal (or near-optimal) plans, rather than a single solution or the first solution found. 
    \end{itemize}
    \item Expressive Numeric Heuristic Search Planner (ENHSP)~\cite{scala2016interval}: planner designed to handle numeric and temporal planning domains.
\end{itemize}

We experiment with two network architectures. The first, featuring two subnets, is described in the motivating example of Section~\ref{sec:motivation_threat:example} and is referred to as \networkTwo in Section~\ref{sec:evaluation:planning}.
The second is based on the Purdue Model~\cite{9471765}, features six subnets and three configurations, and is described in Section~\ref{sec:evaluation:six}.

\subsubsection{Purdue model architecture}
\label{sec:evaluation:six}
\begin{figure}[ht]
    \centering
    \includegraphics[width=\linewidth]{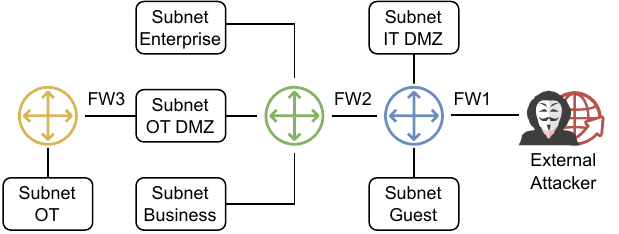}
    \caption{Architecture based on the Purdue model}
    \label{fig:six_subs}
\end{figure}

Figure~\ref{fig:six_subs} provides an overview of the architecture based on the Purdue Model. 

The network is organized into six distinct subnets, each serving a specific purpose. IT DMZ and Guest subnets represent Level 5 in the Purdue Model, which contains all the hosts that can be exposed on the Internet. The Enterprise and Business subnets represent Level 4 in the Purdue Model, which contains general-purpose network devices (e.g., employee devices, internal servers, databases). The OT DMZ subnet is the point of contact between the enterprise and industrial networks. Finally, the OT subnet contains all the industrial devices that manage the industrial processes.

Each subnet is specialized as follows:
\begin{itemize}
    \item IT DMZ: contains the servers accessible from the Internet, providing services to external users.
    \item Guest: includes devices designated for use by non-employees, such as temporary visitors.
    \item Enterprise: contains servers dedicated to IT infrastructure management and administrative functions.
    \item Business: contains servers that store and manage critical information (e.g., database servers).
    \item OT DMZ: contains all the hosts that make possible the interaction between the Enterprise and the Business network. 
    \item OT: contains Supervisory Control and Data Acquisition (SCADA) systems and the essential components for monitoring and controlling industrial processes (e.g., PLC, IED).
\end{itemize}

The connectivity is managed by three routers ($FW1$, $FW2$, $FW3$) which also enforce firewall rules:
\begin{itemize}
    \item \textbf{$FW1$}: connects the IT DMZ subnet, Guest subnet, $FW2$, and the link that provides internet access. It also establishes a trusted communication channel between $web\_server2$ and $FW2$.
    \item \textbf{$FW2$}: serves as the central router connecting $FW1$, the OT DMZ, the Enterprise subnet, and the Business subnet.
    \item \textbf{$FW3$}: Links the OT DMZ hosts to the OT subnet. 
\end{itemize}

To evaluate scalability and generalization, we configured three variations of the network architecture: \purdueOne, a base configuration with 20 hosts; \purdueTwo, a scaled version in which the original hosts are replicated to create a larger network of 200 hosts, simulating a more realistic environment; and \purdueThree, a 200-host variant with increased heterogeneity in the product stack, designed to increase the number of exploits (and thus planning actions) in the PDDL domain file, potentially increasing the number and diversity of exploit chains discovered by \name and the execution time.
The host scaling in configuration \purdueTwo{ }primarily increases the size of the PDDL problem file by expanding the search space size represented by the number of network entities and potential targets. In contrast, configuration \purdueThree{ }increases the number of distinct exploits (i.e., planning actions), thereby enlarging the PDDL domain file.

Table~\ref{tab:hosts_of_20_example_and_techstack} details the product stack for each host in the \purdueOne setup.
The starting point of the attacker is external to the network, while the target host ($\mathbf{scada4}$) is located within the \textit{OT} subnet.

\begin{table*}[ht]
    \centering
    \caption{Application stacks, operating systems, and host assignments across subnets in the \purdueOne network. The target host~(scada4) is highlighted in \textbf{bold}.}
    \begin{tabular}{l|l|lp{6.2cm}}
        \textbf{Subnet} & \textbf{Host} & \textbf{OS} & \textbf{Vulnerable application stack} \\\hline
        IT DMZ & $web\_server1\textnormal{--}3$ & Debian Linux 10.0.0 & Apache Log4j 2.14.0\\\hline
        Guest & $guest1$ & Windows & GOG Galaxy 2.0.12, Mozilla Firefox 10.0.12 \\
        Guest & $guest2$ & Android 4.1.2 & Android API 16.0.0 \\\hline
        Enterprise & $corp1\textnormal{--}2$  & Linux Kernel 4.12.6 & FreeBSD 1.0.0, AnyDesk 5.5.2, VMWare Player 5.0.2\\\hline
        Business & $data1\textnormal{--}3$ & Linux Kernel 4.12.6 & Apache CouchDB 2.0.0\\\hline
        OT DMZ  & $lan1$ & Linux & runc 1.1.11, Elastic Kibana 6.6.0\\
        OT DMZ  & $lan2\textnormal{--}3$ & Unspecified & Elastic Kibana 6.6.0\\
        OT DMZ  & $lan4\textnormal{--}5$ & Windows Server 2012 & Unspecified \\\hline
        OT & $scada1\textnormal{--}2$ & Windows & Berlios GPS Daemon 2.7.0 \\
        OT & $scada3$ & Windows & 7T IGSS 1.0.0 \\
        OT & $\mathbf{scada4}\textnormal{--}5$ & Windows & CitecSCADA 7.0.0 \\

    \end{tabular}
    \label{tab:hosts_of_20_example_and_techstack}
\end{table*}



\subsection{AI Planning Module Capabilities}\label{sec:evaluation:planning}

Section~\ref{sec:discovery-multiple-plans} shows how \name discovers multiple exploit chains as the network size increases, Section~\ref{sec:different_sources_different_targets} shows \name{'} generalizability by varying the exploit source and changing the target host, Section~\ref{sec:pddl-prune-plan} assesses planning efficiency and scalability, and Section~\ref{sec:analysis-privilege-levels} presents a sensitivity analysis over privilege levels.

\subsubsection{Capacity to Discover Multiple Exploit Chains on Larger networks}\label{sec:discovery-multiple-plans}

To determine whether \name can uncover multiple exploit chains in the same environment, we focus on the K* planner’s ability to return more than one valid plan.
Specifically, we measure, on each network described in Section~\ref{sec:evaluation:experimental_setup}, how many distinct plans K* produces, the average plan length (i.e., mean number of exploits per chain), and the computation time required for their identification.

Table~\ref{tab:k_planning} shows the performance of K* when a single attacker/target pair is fixed while we scale both the number of hosts in the network and the number of possible actions (i.e., exploits).
The results highlight significant increases in search time as the network complexity grows. Specifically, the runtime scales by an order of magnitude between \networkTwo and \purdueOne, and by two further orders of magnitude between \purdueOne{ }and \purdueTwo, with 83 actions. An additional order of magnitude increase is observed when scaling from 83 to 114 actions in \purdueThree. 
However, even the largest scenario (200 hosts) is completed in just 26.258 seconds, keeping the planning task in a practical range.
K* consistently uncovers 13 unique plans in every topology, with an average chain length of 3.6 exploits for \networkTwo and 6.9 exploits for the other topologies.
These findings suggest that while K* successfully identifies multiple unique plans, it is accompanied by increased chain length and planning time as network size and action set complexity increase.".
Regardless, the planner's duration remains within acceptable limits even in more complex settings and running on commodity hardware, suggesting its potential for handling larger problem instances, albeit with an increase in computational cost.

\begin{table}[tb]
    \centering
    \caption{Target network, number of hosts, number of actions, number of unique plans found by  K*, average chain length (c), and planning duration in seconds~(10 trials).}
    \begin{tabular}{lcc|ll}
         \textbf{Network} & \textbf{\# Hosts} & \textbf{\# Actions} & \textbf{\# Plans} & \textbf{Duration (s) (Std)} \\
         \hline
            \networkTwo & 2 & 20 & 13 (c: 3.6) & 0.008 (0.000) \\
            \purdueOne & 20 & 83 & 13 (c: 6.9) & 0.016 (0.001) \\
            \purdueTwo & 200 & 83 & 13 (c: 6.9) & 7.556 (0.127) \\
            \purdueThree & 200 &114 & 13 (c: 6.9) & 26.258 (5.846) \\
    \end{tabular}
    \label{tab:k_planning}
\end{table}

\subsubsection{Different Exploit Sources, Different Targets}\label{sec:different_sources_different_targets}
In realistic security assessment scenarios, testers must assume that different hosts could be the target of the attacker.
We therefore ran K* on \networkTwo and \purdueOne{ }once per host, and repeated the experiment with two independent exploit corpora (Metasploit and \coreblah~\cite{coresec}) to verify that our workflow is not tied to a specific tool set.
Table~\ref{tab:different_host} presents the number of plans found for each host in \networkTwo and \purdueOne, while varying the exploit source.

Using Metasploit on the small \networkTwo, \name discovers a total of 21 exploit chains: 8 targeting \hostone and 13 targeting \hosttwo. 
The larger \purdueOne{ }network exhibits far richer attack surfaces: every data host (i.e., $data1\textnormal{--}3$) can be compromised via 30 chains. At the same time, hosts in the OT DMZ (i.e., $lan4\textnormal{--}5$) and OT (i.e., $scada3\textnormal{--}5$) subnets can be compromised with 13 chains. All nodes in the IT DMZ subnet (i.e., $web\_server1\textnormal{--}3$) can be compromised as well, with \textit{$web\_server2$} slightly more exposed (18 chains) than its peers.

Replacing Metasploit modules with \coreblah exploits reshapes the landscape. \networkTwo now yields 29 total chains with 5 chains targeting \hostone and 24 targeting \hosttwo. 
On the other hand, the number of plans on \purdueOne{ }drops to 102 (a decrement of 50\%). The drop is concentrated on the nodes on the IT DMZ subnet, each now at three chains, and on the nodes on the OT DMZ subnet, which cannot be compromised using \coreblah exploits. 

\begin{table}[tb]
    \centering
    \caption{Impact of changing the target host. The number of exploit chains discovered by K* when targeting every host in \networkTwo and \purdueOne{ }using Metasploit and \coreblah (Core) exploits.}
    \begin{tabular}{l|l|cc}
        &&\textbf{Metasploit} & \textbf{Core}\\
         \textbf{Network} & \textbf{Host} & \textbf{Plans} & \textbf{Plans}\\\hline
           \multirow{2}{*}{\networkTwo} & \hostone & 8 & 5\\
           & \hosttwo & 13 & 24\\\hline
          \networkTwo & \textbf{TOTAL} & \textbf{21} & \textbf{29}\\\hline
           \multirow{11}{*}{\purdueOne}& $data1$ & 30 & 21 \\
           & $data2$ & 30 & 22 \\
           & $data3$ & 30 & 20 \\
           & $lan4$ & 13 & 15 \\
           & $lan5$ & 13 & 15 \\
           & $scada3$ & 13 & 0\\
           & $scada4$ & 13 & 0\\
           & $scada5$ & 13 & 0\\
           & $web\_server1$ & 16 & 3\\
           & $web\_server2$ & 18 & 3\\
           & $web\_server3$ & 16 & 3\\\hline
          \purdueOne & \textbf{TOTAL} & \textbf{205} & \textbf{102} \\
         \end{tabular}
    \label{tab:different_host}
\end{table}

\subsubsection{Different Planners and Their Performance Capabilities}\label{sec:pddl-prune-plan}

To evaluate the efficiency and scalability of \name, we consider the time to find one exploit chain when varying the off-the-shelf AI planner and network. We measure the time required to identify a valid plan within each network. 
This provides insights into the scalability of each planner with increasing network complexity, considering factors such as the number of subnets, hosts, and exploits referenced in the exploit source.

Table~\ref{tab:results_planner_pddl_variants}  reports mean runtimes in seconds after 10 trials while varying both the number of hosts in the PDDL problem file and the number of exploits/actions in the PDDL domain file for each planner. The right-most column reports the duration required to compute a single plan, with standard deviations noted in parentheses. 

In \networkTwo, all planners solve the instance in sub‑second time, but K* (0.002 seconds) is an order of magnitude faster than the others; ENHSP follows at 0.07 seconds, while LAMA and LAMA‑first yield results around 0.3 seconds.

In \purdueOne, scaling both the number of hosts to 20 and the number of actions to 83, K* maintains its lead with a runtime of 0.01 seconds, showcasing robust scalability. LAMA‑first and ENHSP remain usable (0.43 and 0.30 seconds, respectively), whereas LAMA slows to 1 second.

Fixing the number of actions, but scaling up the number of hosts in \purdueTwo{ }reveals significant performance degradation for some planners.
LAMA exceeds 40 minutes, becoming impractical for larger networks, and ENHSP increases its runtime to 51.61 seconds. 
In contrast, K* (3.25 seconds) and LAMA‑first (4.48 seconds) stay within interactive limits, highlighting stronger pruning heuristics.

Adding 31 extra actions in \purdueThree{ }barely impacts LAMA‑first, which increases its runtime by only 1.3 seconds; K* and ENHSP show no significant slowdown relative to the previous case (no statistical difference).

Across all configurations, K* exhibits the shortest runtimes as both the problem (host count) and domain (exploit count) grow, underscoring its superior scalability and efficiency. These results indicate that K* is the most suitable off‑the‑shelf planner for \name.
LAMA-first is a viable alternative for larger networks, though it is slower than K* and cannot be queried for multiple plans. ENHSP is competitive on small instances yet suffers in memory‑intensive scenarios, while LAMA becomes impractical once the problem space exceeds hundreds of hosts. These results confirm that \name remains usable on networks a hundred times larger than the motivating example.

\begin{table}[tb]
    \centering
    \caption{Average duration (in seconds) over 10 trials for a planner when queried for a single plan on various example networks.}
    \begin{tabular}{l|c|c|l|l}
        \multicolumn{3}{c}{\textbf{Example Network}} & \multicolumn{1}{c}{\textbf{}} & \textbf{Perf.} \\        
         \textbf{Network} & \textbf{\# Hosts} & \textbf{\# Actions} & \textbf{Planner} & \textbf{Duration (s)} \\
         \hline
          \networkTwo & 2 & 20  & LAMA-first & 0.30 (0.01) \\
          \networkTwo & 2 & 20  & LAMA & 0.31 (0.02) \\
          \networkTwo & 2 & 20  & K*  & 0.002 (0.000) \\
          \networkTwo & 2 & 20  & ENHSP & 0.07 (0.00) \\ \hline
          \purdueOne & 20 & 83  & LAMA-first & 0.43 (0.02) \\
          \purdueOne & 20 & 83  & LAMA & 1.06 (0.01) \\
          \purdueOne & 20 & 83  & K* & 0.01 (0.000) \\
          \purdueOne & 20 & 83  & ENHSP & 0.30 (0.01) \\  \hline
          \purdueTwo & 200 & 83  & LAMA-first & 4.48 (0.02) \\
          \purdueTwo & 200 & 83  & LAMA & 2,574 (450) \\
          \purdueTwo & 200 & 83  & K* & 3.25 (0.07) \\
          \purdueTwo & 200 & 83  & ENHSP & 51.61 (5.67) \\ \hline
          \purdueThree & 200 & 114  & LAMA-first & 5.77 (0.04) \\
          \purdueThree & 200 & 114  & K* & 3.16 (0.06) \\
          \purdueThree & 200 & 114  & ENHSP & 46.62 (6.47) \\
    \end{tabular}
    \label{tab:results_planner_pddl_variants}
\end{table}

\subsubsection{Misconfigured Privilege Levels}
\label{sec:analysis-privilege-levels}
Incorrectly assigning operating privileges to network services is a well‑known root cause of successful intrusions. 
This impacts whether the service is exploitable and is particularly dangerous when services are assigned excessively high privileges~\cite{gu2024epscan}. 
To quantify how privilege misconfigurations influence \name{’}s output, we conduct a sensitivity analysis that artificially shifts the acquired privilege of most of the \RCE exploit in the domain.
We established three scenarios:
\begin{itemize}
    \item Baseline: privileges remain as predicted during the Classification phase.
    \item Upper Bound (\textbf{UB}): \RCE exploits with low acquired privileges are forcibly granted high privileges. 
    \item Lower Bound (\textbf{LB}): \RCE exploits with high acquired privileges are forcibly granted low privileges.
\end{itemize}

For each scenario, we run the K* planner and count the distinct exploit‑chain plans returned.
This allows us to estimate the implications of the misconfigurations in terms of the number of exploit chains. 

The results in Table~\ref{tab:privileges} demonstrate that, in the UB scenario, the set of plans discovered is unchanged. 
By contrast, in the LB scenario, \name loses 2 of 13 chains in the \networkTwo and 6 of 13 chains in each larger Purdue‑style network. 
The reduction is modest but significant, indicating that a correct privileges configuration does not eliminate the risk of finding chains. Even in the LB scenario, every network still contains seven exploit chains, confirming that some exploit chains are strictly dependent on the attack surface created by product combinations and topology, regardless of whether a pessimistic or optimistic configuration is used.

\begin{table}[tb]
    \centering
    \caption{Impact of misconfigured privilege levels. Number of plans found if every application is configured with all low privileges (LB), if every application is configured with all high privileges (UB), and the baseline with the acquired privileges manually labeled and predicted by GPT-4o.}
    \begin{tabular}{lc|c|c|c}
        \textbf{Network}& \textbf{\# Actions} & \textbf{LB} & \textbf{UB} & \textbf{Baseline}\\\hline
        \networkTwo & 20  & 11 & 13 & 13 \\
        \purdueOne & 83 & 7 & 13 & 13\\
        \purdueTwo & 83 & 7 & 13 & 13 \\
        \purdueThree & 114 & 7 & 13 & 13 \\
    \end{tabular}
    \label{tab:privileges}
\end{table}

\subsection{Analysis of the Exploit Classification Module}\label{sec:evaluation:class}
Because \name relies on an LLM to classify exploits before they are fed into the Modeling step, it is crucial to understand how accurately the LLM performs this task. 
We evaluated GPT-4o's classification capabilities by verifying its performance on a representative sample of 100 exploits from Metasploit. 
Fifty exploits were drawn uniformly at random; the remaining fifty were selected with k‑means over sentence‑transformer embeddings to ensure coverage of diverse patterns within the dataset. The final corpus contains 90 \RCE exploits and 10 PE exploits.
For every exploit, to set up the ground truth, we manually labeled the four attributes used during the modeling step (i.e., Exploit Type, Protocol, Privileges Required, Privileges Acquired). GPT‑4o was then queried with the same prompt employed in the automated pipeline, and its predictions were compared to the ground‑truth labels. Class‑weighted precision, recall, and F1 scores were computed for each attribute as well as for their joint prediction (Overall) to compensate for class imbalance in the dataset.

Table~\ref{tab:llm_exploit_acc} shows the classification recall, precision, and F1 Score of GPT-4o for Exploit Type (PE or \RCE), \RCE Protocol (TCP or UDP), Privileges Required (N, L, H), and Privileges Acquired (L, H, R).
The results for the LLM are highly promising, particularly for Exploit Type (F1 Score: 0.96), \RCE Protocol (F1 Score: 0.95), and Privileges Required (F1 Score: 0.93), showing that GPT-4o excels in classifying the coarse‑grained features. However, lower scores were observed for Privileges Acquired (F1 Score: 0.75), which in turn impacted the Overall result (F1 Score: 0.71).
This is unsurprising: some privileges acquired depend on run‑time configuration rather than the exploit itself. Consequently, mis‑labeling privileges acquired occasionally causes \name to miss potential exploit chains or to result in exploit chains that could fail because some exploit in the chain assumes and relies upon the overestimation. However, despite these limitations, \name consistently discovers viable exploit chains across all tested scenarios. 

\begin{table}[tb]
    \centering
    \caption{Classification results on 100 hand-labeled exploits representative of Metasploit, \textbf{bold} marks highest and \textit{italics} marks lowest. Results are split into the different components of our classes: Exploit Type (\RCE or PE), Protocol (TCP or UDP; for \RCE only), Privileges Required (N, L, or H), and Privileges Acquired (L, H, or R). The reported Precisions, Recalls, and F1 Scores are weighted by the distribution of the different labels in our hand-labeled set. The average of 10 runs and the standard deviation are shown. Note that weighted recall is equivalent to accuracy.}
    \begin{tabular}{l|lll}
               & \textbf{Recall} & \textbf{Precision} & \textbf{F1} \\ \hline
    Exploit Type       & \textbf{0.96} (0.01)    & \textbf{0.97} (0.01)      & \textbf{0.96} (0.01)     \\
    Protocol   & 0.94 (0.01)   & 0.95 (0.01)      & 0.95 (0.01)     \\
    Priv. Req. & 0.92 (0.01)   & 0.94 (0.01)      & 0.93 (0.01)     \\
    Priv. Acq.  & \textit{0.75} (0.03)    & \textit{0.76} (0.02)      & \textit{0.75} (0.03)     \\
    \hline
    Overall        & 0.69 (0.03)    & 0.75 (0.01)      & 0.71 (0.02)    
    \end{tabular}
    \label{tab:llm_exploit_acc}
\end{table}




\section{Discussion and Future Work}\label{sec:discussion}
We demonstrated effectiveness and efficiency in generating and identifying multiple exploit chains, adapting seamlessly to alternative exploit databases.
For the underlying AI planning step in \name, the K* planner consistently outperformed alternatives in terms of runtime and scalability, particularly in larger network configurations. Its ability to efficiently compute both single and multiple plans demonstrates robustness across a range of scenarios, which is crucial for real-world applications in complex network environments. Moreover, scaling challenges can be mitigated by dividing the network into parts, allowing each to be analyzed independently without compromising overall coverage.

Privilege configurations were also found to play a critical role in determining the feasibility of exploit chains, however, some exploit chains are intrinsically tied to the product combinations and network topology. Moreover, since downstream tasks such as risk scoring and impact assessment rely on accurate chain modeling, improving privilege inference remains an important direction for reducing uncertainty in risk calculation metrics.

While we demonstrated promising results with \name, there remain certain limitations and opportunities for further improvements.

Firstly, the effectiveness of \name in identifying valid exploit chains relies on the availability of correct exploit labels, including well-defined connections to vulnerable configurations. While GPT-4o can generate these labels with good precision, recall, and F1 Score, there is potential for further improvement by incorporating higher-quality data (e.g., from private sources).

Secondly, the success of exploit chain discovery is highly dependent on the completeness and correctness of the network modeling. Incomplete or ambiguous descriptions of network components and their interconnections may hinder the effective discovery of exploit chains. Future work could integrate network scanning tools that extract host and service information, generate network topology graphs, and assist in producing problem files, thereby reducing errors from manual modeling and automating the entire process.

Moreover, to fully confirm an exploit chain, the exploit code must be functional. Currently, \name has a semi-automated method for executing exploit chains. Future work would be the development of an automated framework to streamline this confirmation process and reduce the reliance on manual intervention.

Finally, future work also involves the development of a data analysis module designed to generate and visualize synthetic risk profiles for network hosts. This module would assign a quantifiable cost to each identified exploit, enabling the prioritization of multiple exploit chains and the identification of high-risk or bottleneck devices within the network.
These risk metrics could also be integrated into the PDDL representations to guide strategic planning and mitigation.

Despite these limitations, \name provides a solid foundation for defenders to assess potential exploit chains within their networks, enabling more informed security decisions.



\section{Related Work}\label{sec:related}
\label{sec:related_work}

\begin{table*}[ht!]
    \centering
    \caption{Comparison between us and the closest related work on key properties.}
    \begin{tabular}{p{3cm}|p{2.8cm}p{2.8cm}p{2.8cm}p{4cm}}
        \textbf{Property/Related Work} & \cite{obes2013attack} (2013) & \cite{9138852} (2020) & \cite{299659} (2024) & Ours \\\hline
        \textbf{Planning Method} & Classical planner & Rule-based attack graph generator & LLM-assisted planner & LLM-assisted planner \\ 
        \hline
        \textbf{Planners}  & Metric-FF, SGPlan & MulVAL & PowerLifted   & K*, LAMA, LAMA-first, ENHSP \\ 
        \hline
        \textbf{Target}         & Networks & Small networks & Single-hosts & Networks \\
        \hline
        \textbf{OS}     & All & All & Unix & All \\
        \hline
        \textbf{Exploit Sources}    & CoreImpact & Metasploit & GTFOBins & \coreblah, Metasploit \\
        \hline
        \textbf{Objective achieved}  & Lateral movement & Lateral movement + PE & PE                  & Lateral movement + PE            \\
    \end{tabular}
    \label{tab:diff}
\end{table*}

Identifying exploit chains has become an important focus in cybersecurity research~\cite{238580} because such chains represent the interconnected steps attackers may take in multi-phase intrusions, revealing critical vulnerabilities that must be addressed to prevent system compromise.
Various methods, models, and tools have been developed to analyze~\cite{ZENITANI2023103081} and generate attack graphs~\cite{KONSTA2024103602} and exploit chains~\cite{10490046}, enabling security professionals to defend against sophisticated, multi-step attacks. 
These models are pivotal in the latest research on key topics such as network hardening~\cite{6263942, 10.1145/3576842.3582326}, moving target defense~\cite{9063635}, and IoT security~\cite{9486846, 8720257}, and are continuously evolving to consider, for instance, insider threats~\cite{DAMBROSIO2023103410} and cyber-physical actions~\cite{BARRERE2023103348,DAMBROSIO2025104315}. Automated tools like MulVAL~\cite{ou2005mulval} and its extensions~\cite{9277665} can generate these structures by analyzing vulnerabilities within a network. This type of tool is also used in frameworks for automated vulnerability assessment through executable attack graphs, which improve the efficiency of identifying exploitable paths within networks~\cite{9138852}. However, existing approaches often struggle to scale effectively when applied to larger and more complex network topologies, due to the combinatorial explosion of possible attack paths and increased computational overhead.

AI planning techniques have been applied in this domain~\cite{obes2013attack, hoffmann2015simulated, hemberg2021using}, offering approaches for automating the generation of attack scenarios~\cite{amos2017efficient}, courses of actions~\cite{10.5555/3037062.3037065}, and mitigation strategies~\cite{speicher2019towards, choi2021plan2defend}. AI planning techniques, such as those based on PDDL, allow attack scenarios to be modeled as a series of actions~\cite{obes2013attack}. 
Recently, LLMs have been employed to extract information from natural language sources like the CVE database~\cite{10628694} and automate the creation of PDDL models~\cite{silver2024generalized,Oswald_Srinivas_Kokel_Lee_Katz_Sohrabi_2024}. By converting unstructured text into structured planning problems, LLMs enable efficient problem formulation and analysis~\cite{liu2023llm+,katz2024thought}.

Recent advances in NLP could enable more use of AI planning in exploit chain discovery, which automates the extraction of security data from text-based sources to translate it into PDDL~\cite{silver2024generalized}. However, research in this area remains limited, with current efforts focusing primarily on the automated generation of PDDL Problem files for single Unix instances~\cite{299659}, dealing only with PE exploits. 

This approach has shown promise in identifying previously unknown or complex exploit chains, offering a scalable solution for vulnerability management and network defense. 
Our research aims to address these limitations by automating the generation of domain files and incorporating \RCE exploits to address more complex network environments, not just single Unix instances.
Table~\ref{tab:diff} highlights the key differences between our approach and the most closely related methods in the literature.

\section{Conclusions}\label{sec:conclusion}
In this paper, we introduced \name, a novel method for discovering chains of PE and \RCE exploits within networks. \name's methodology consists of three key steps: (1) classifying exploits from a data source, (2) modeling the network and exploits as PDDL problem and domain files, and (3) employing an AI planner to identify exploit chains that can be executed during manual penetration testing.

We validated \name using a motivating example involving a firewalled network with two hosts configured with realistic technology stacks and three planted vulnerabilities that could be chained together. While \name successfully uncovered the expected exploit chain, it also discovered twelve unanticipated chains, one of which we manually executed.

\name demonstrates both speed and scalability, detecting multiple exploit chains in networks with up to 200 hosts in under 30 seconds. The method proves effective across various network architectures and configurations, utilizing different exploit data sources (Metasploit and \coreblah). Notably, it is able to discover multiple exploit chains, even with minimal privileges. For demonstration purposes, we implemented \name with Metasploit, a widely available and public penetration-testing framework, and also integrated it with \coreblah.

The classification results for \name are promising, particularly in the areas of Exploit Type, \RCE Protocol, and Privileges Required, with high F1 scores of 0.96, 0.95, and 0.93, respectively. However, the classification of Privileges Acquired showed a lower F1 score of 0.75, mainly due to the challenges in inferring specific application privilege configurations with only public data. 
Regarding the planning step, the use of the FD K* planner demonstrated both efficiency and scalability. It identified an exploit chain in just 0.002 seconds in a smaller network, while also scaling effectively to larger networks, finding an exploit chain in 3.16 seconds.

In conclusion, \name offers a powerful methodology for defenders to assess potential exploit chains within their networks, helping to enhance security posture and facilitate more informed defense strategies.


\newpage




%
\bibliographystyle{IEEEtran}
\bibliography{bibliography}

\begin{appendices}
\section{Exploit Classification prompt}
The exploit classification prompt is structured in three parts:
\begin{itemize}
    \item \textbf{System Prompt}: This section outlines the LLM’s task, specifying the classification objective, privilege levels available with keywords to aid in identifying key exploit characteristics (Listing~\ref{listpriv}), and classes of our network model with brief descriptions (Listing~\ref{listclass}).
    \item \textbf{Few-shot Examples} (Listing~\ref{lstprivesc}): Six few-shot examples are provided to demonstrate the format and expectations for classification. Each example contains the Metasploit description, CVE ID and description, the vulnerable configuration of CPEs, and the corresponding classification result.
   \item \textbf{Target Exploit}: In the final section, we present the details of the specific exploit to be classified like in the few-shot examples. 
\end{itemize}

\begin{lstlisting}[label=listpriv,caption=System prompt - privileges description,float,frame=tb, backgroundcolor = \color{lightlightgray},captionpos=b]
System: You are a tool for cybersecurity experts. When given
a description of an exploit, you should classify it into one of the
following classes. Use the descriptions of each class to help guide you.
ONLY return the name of the class with no other text and no explanation.

Here is the definition of each type of privilege and keyword to help guide you in your classification.
    no privileges: Can only send network traffic to listening services on the host. Unauthenticated user.
        - unauthenticated
        - no privileges
        - unauthenticated user
    low privileges: Authenticated user with non-admin privileges.
        - user level
        - non-admin
        - authenticated
    high privileges: Authenticated user with admin level privileges. Almost all services on a remote host can be assumed to be running with high privileges.
        - admin
        - arbitrary code execution
        - authenticated
    root privileges: Privileged user with root or SYSTEM level privileges. If no mention of root or SYSTEM level privileges are present, do not assume root privileges.
        - root
        - SYSTEM
\end{lstlisting}

\begin{lstlisting}[label=listclass,caption=System prompt - classes description,float,frame=tb, backgroundcolor = \color{lightlightgray},captionpos=b]
PE_L_H: A privilege escalation exploit that requires low privileges and obtains high privileges. This exploit runs locally on a single host.
...
PE_H_R: A privilege escalation exploit that requires high privileges and obtains root privileges. This exploit runs locally on a single host.
RCE_TCP_N_L: A remote code execution exploit over the TCP protocol. Given an agent with no privileges on a remote host, obtains low privileges on the remote host.
...
RCE_TCP_H_R: A remote code execution exploit over the TCP protocol. Given an agent with high privileges on a remote host, obtains root privileges on the remote host.
RCE_UDP_N_L: A remote code execution exploit over the UDP protocol. Given an agent with no privileges on a remote host, obtains low privileges on the remote host.
...
RCE_UDP_H_R: A remote code execution exploit over the UDP protocol. Given an agent with high privileges on a remote host, obtains root privileges on the remote host.
\end{lstlisting}

\begin{lstlisting}[label=lstprivesc,caption= Few-shot - PE example,float,frame=tb, backgroundcolor = \color{lightlightgray},captionpos=b]
Human: Exploit Description:
This module exploits a command injection vulnerability in IBM AIX invscout set-uid root utility present in AIX 7.2 and earlier. The undocumented -rpm argument can be used to install an RPM file; and the undocumented -o argument passes arguments to the rpm utility without validation, leading to command injection with effective-uid root privileges. This module has been tested successfully on AIX 7.2.

CVE Descriptions:
CVE-2023-28528
IBM AIX 7.1, 7.2, 7.3, and VIOS 3.1 could allow a non-privileged local user to exploit a vulnerability in the invscout command to execute arbitrary commands.  IBM X-Force ID:  251207.

Vulnerable Versions:
[[{'children': [], 'cpe_match': [{'cpe23Uri': 'cpe:2.3:o:ibm:aix:7.1:*:*:*:*:*:*:*', 'cpe_name': [], 'vulnerable': True}, ..., {'cpe23Uri': 'cpe:2.3:o:ibm:aix:7.3:*:*:*:*:*:*:*', 'cpe_name': [], 'vulnerable': True}], 'operator': 'OR'}]]
AI: PE_L_R
\end{lstlisting}
\end{appendices}

\end{document}